\documentclass[12pt]{article}
\usepackage{graphicx}

\newcommand\un[1]{{\,\rm #1}}
\newcommand\E[1]{\times10^{#1}}
\newcommand\rs[1]{_\mathrm{#1}}

\newcommand\apj{ApJ}
\newcommand\apjl{ApJL}
\newcommand\aap{A\&A}
\newcommand\mnras{MNRAS}

\begin{document} 

   \textbf{\LARGE Three-dimensional velocity fields in the silicon- and sulfur-reach ejecta in the remnant of Tycho supernova}\\
   
   \textit{O. Petruk$^{1,2}$,
          M. Patrii$^{3}$,
          T. Kuzyo$^{1}$,
          A. Baldyniuk$^{3}$,
          V. Marchenko$^{4}$,
          V. Beshley$^{1}$
          }\\

   $^1$Institute for Applied Problems in Mechanics and Mathematics, National Acade\-my of Sciences of Ukraine, Naukova 3-b, 79060 Lviv, Ukraine

        $^2$INAF - Osservatorio Astronomico di Palermo, Piazza del Parlamento 1, 90134 Palermo, Italy
        
        $^3$Faculty of Physics, Ivan Franko National University of Lviv, Kyryla and Methodia  8, 79005 Lviv, Ukraine
        
        $^4$Astronomical Observatory, Jagiellonian University, Orla 171, 30-244 Cracow, Poland\\

  \textit{The three-dimensional {velocity structure of the shock-heated Si-reach and S-reach ejecta} were reconstructed in Tycho supernova remnant {from Doppler-shifted lines}. The vector components along the line of sight were restored from the spatially resolved spectral analysis of the Doppler shifts of Si XIII and S XV lines. The components in the plane of the sky were derived from analysis of the proper motion of the remnant's edge at different azimuths. This has been done by using the data of X-ray observations from Chandra observatory as well as the radio data from the Very Large Array. 
  Differences in Doppler velocities over the Tycho's SNR are of the order of thousands of km/s. The speed of the ejecta on the opposite sides of the remnant as a three-dimensional object differs on $20-30\%$. There are asymmetries and differences in the spatial distributions between the Si-reach and S-reach ejecta components. Namely, the level of isotropy is higher in Si while the vector components directed outward of the observer are larger in S. This puts limitations on the level of deviation of the internal structure of the progenitor star from the ideal layered structure as well as on the level of asymmetries in supernova explosion.}\\


\section{Introduction}

Three-dimensional (3D) models of supernova explosions typically demonstrate a large asymmetry \cite{2005NatPh...1..147W,2008ApJ...681.1448J,2015A&A...577A..48W,2023MNRAS.518.1557R}. Therefore, it would be quite important to look for some observational evidence about the level of anisotropy in a supernova explosion. Some young supernova remnants (SNRs) that exhibit prominent emission from stellar ejecta could be a good target for such a study. Such a remnant keeps this information encoded in the spatial distribution of ejecta velocities. One can decipher the 3D spatial structure of ejecta in two steps. Namely, the Doppler effect may be used for determination of the velocity component along the line of sight, while the proper motion could give us the component in the plane of the sky. 

There are just a few young SNRs with high-resolution observations covering almost all wavelengths of the electromagnetic spectrum, which enable detailed morphological studies. One such object is the Tycho's supernova remnant (G120.1+1.4), the remnant of a supernova explosion of type Ia which happened in our Galaxy and observed in 1572 by Tycho Brahe \cite{2017hsn..book..117D}.

There are studies of the Doppler effect \cite{2017ApJ...842...28W,2017ApJ...840..112S,2023A&A...680A..80G,2024arXiv240417296G,2024ApJ...962..159U} as well as of the proper motion 
\cite{1997ApJ...491..816R,2010ApJ...709.1387K,2013ApJ...770..129W,2021ApJ...906L...3T} in Tycho's SNR. They do indicate an anisotropy in the SNR's shock expansion velocity.
However, the novelty of our task consists in the reconstruction of the 3D velocity fields of ejecta components, which are rich in silicon and sulfur, as well as in the determination of the differences in the spatial distributions of these elements. We have chosen these two species because they are responsible for the two very prominent spectral lines in X-rays and the statistic in the data of X-ray observations is sufficient to perform the spatially resolved spectral analysis.

\section{Observations}

\begin{figure}
  \centering 
  \includegraphics[trim=0 0 40 35, clip,width=7.5truecm]{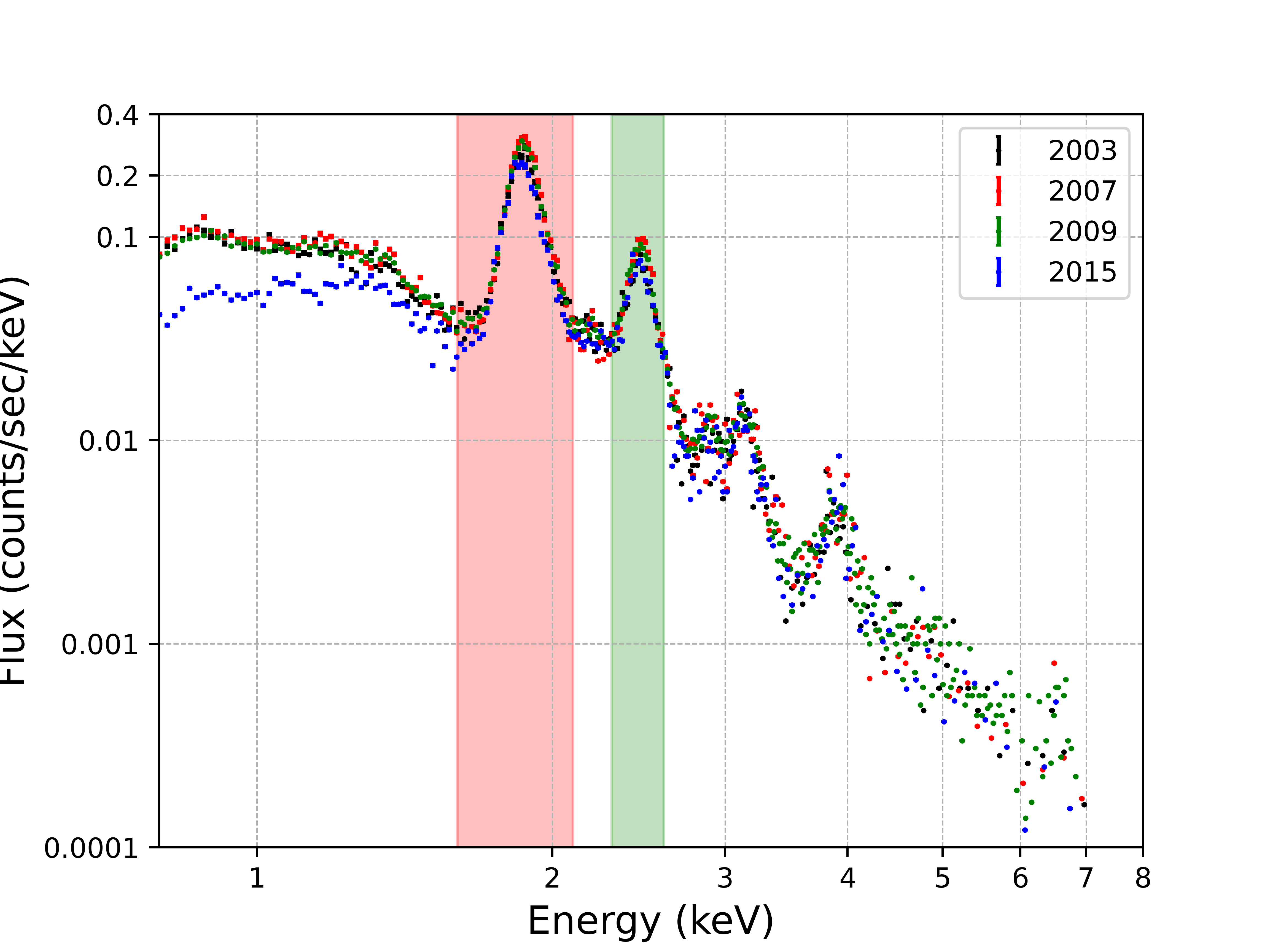}
  \caption{X-ray spectra of Tycho's SNR from four epochs of observations marked by different colors. The red and green shadows outline the photon energies where emission in Si (1.6-2.1 keV) and S (2.3-2.6 keV) lines dominates. The deviation of the blue dots from the others at energies below 1.6 keV is due to the instrument degradation.  
  }
  \label{tycho3d:fig-spectrum}
\end{figure}

\begin{figure}
  \centering 
  \includegraphics[trim=80 20 90 20, clip,width=7.5truecm]{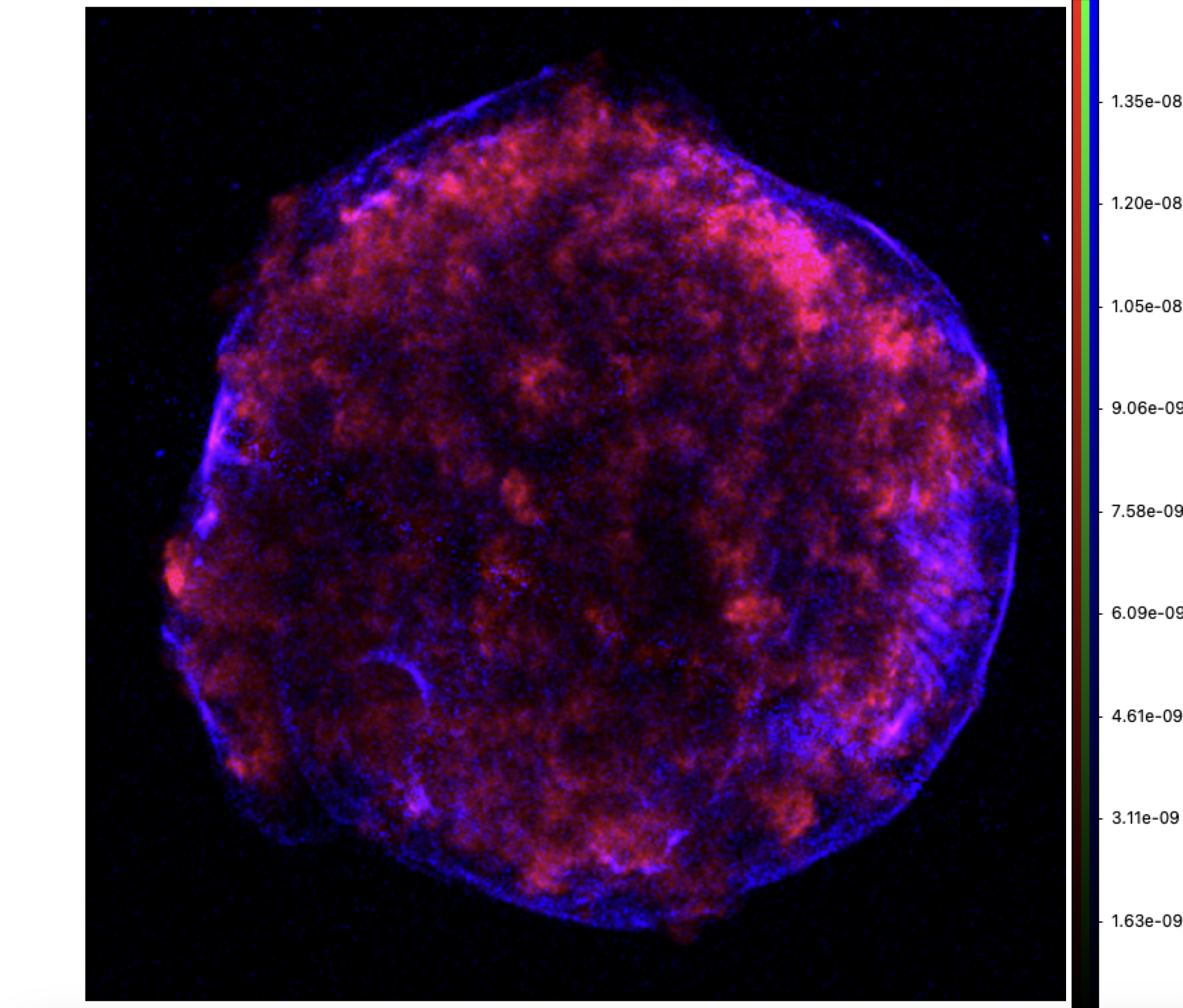}
  \caption{X-ray image of the remnant of Tycho's {supernova} produced from the Chandra data for the year 2015. The red color represents the emission in the photons with energy 1.2-4.0 keV where the thermal emission from the ejecta is dominant. The blue color corresponds to emission in 4.1-6.0 keV where the major contribution comes from the non-thermal radiation of accelerated electrons.
  }
  \label{tycho3d:fig-image}
\end{figure}

The observational data of Tycho's SNR in X-rays were obtained by ACIS-I array of Advanced CCD Imaging Spectrometer (ACIS) on board the Chandra X-ray observatory. The ACIS detector provides the opportunity to simultaneously acquire high-resolution images and moderate resolution spectra. 
The Tycho's SNR observational data were downloaded from Chandra data archive ChaSeR and processed using CIAO-4.12 software with modeling and fitting package Sherpa. 
The imaging application SAOImage ds9 allows to work with astronomical images and visualize data. It supports images in the fits format, allows users to manipulate selected regions, contains various scaling algorithms, and the ability to create multicolored maps.
In order to obtain the combined X-ray spectrum from the entire SNR for each year 2003, 2007, 2009 and 2015 a circular region covering the edges of the remnant was chosen with SAOImage ds9 application. Since the procedure also requires a background region, the background region outside the remnant, which is located next to it, was also chosen. Further, the spectrum was obtained for each observation separately with \verb+specextract+ CIAO command and combined with \verb+combine_spectra+ CIAO command for each selected year to obtain better statistics. 
Fig.~\ref{tycho3d:fig-spectrum} left shows the X-ray spectrum of Tycho's SNR from a number of observations. 
A combined image of Tycho's SNR in X-rays is presented on Fig.~\ref{tycho3d:fig-image} right where the emission in the photon energy range 1.2-4.0 keV, which includes the two most prominent lines of Si and S, is shown in the red color and the hard continuum 4.1-6.0 keV in blue color. 

In our work, we also use the radio maps of the remnant of Tycho's SNR at 1.4~GHz  derived from observations performed on the radio interferometer Very Large Array (VLA) and published in \cite{2016ApJ...823L..32W}.


\section{Dopplerography}
\label{tycho3Da:sect-doppl}

In the present section, we would like to derive the spatial distributions of the Doppler shifts and velocities for the two brightest lines in the X-ray spectrum of Tycho's SNR, namely, for Si ($1.6-2.1\un{keV}$) and S ($2.3-2.6\un{keV}$).

For the spatially resolved spectral analysis, we use the Chandra data derived in observations during the year 2009, which have the longest integrated exposure (Table~\ref{tycho3d:table-observ}). We created a mesh of $20\times20$ square cells covering the Tycho's SNR. 
The number of counts observed in each cell over the SNR is above $3\E{4}$ for $1.6-2.1\un{keV}$, $1\E{4}$ for $2.3-2.6\un{keV}$ and $1500$ for $4.1-6\un{keV}$. Such high numbers of events provide low errors in the line shapes and serve for accurate fitting.

\begin{figure}
  \centering 
  \includegraphics[width=7.5truecm]{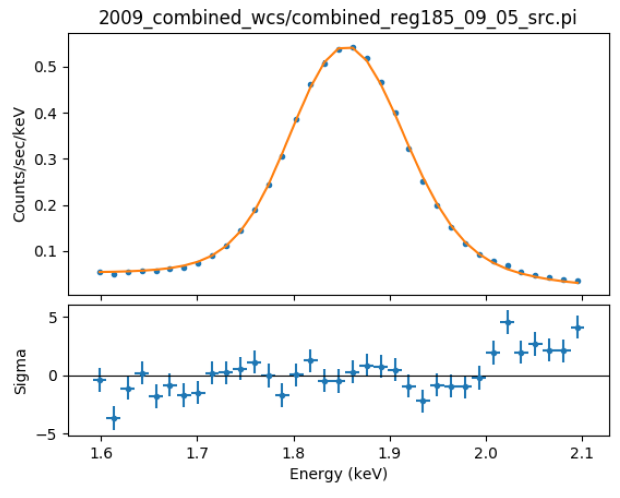}
  \includegraphics[width=7.5truecm]{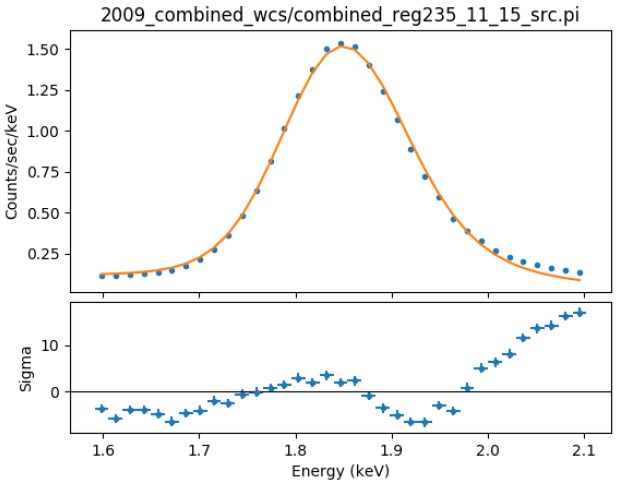}
  \caption{
    Examples of shapes of Si lines in two cells, with a `good' (\textit{top}, 
    $rstat=3.32$) and `worse' (\textit{bottom}, 
    $rstat=51.4$) fits. The number of such `worse' fits is minor. Most of them were improved by manual selection of parameters for the \textit{gauss1d} model and/or by narrowing the photon energy range for the fit. For example, the manual fit results in $rstat=1.71$ for the case shown on the bottom plot.
  }
  \label{tycho3d:fig-linefits}
\end{figure}

The procedure we perform for each cell over the SNR is the following. First, we fit the spectrum at high energies ($4.1-6\un{keV}$) by the pure continuum model (\textit{xsbremss}). {Though the non-thermal emission contributes to the emission in this band, we limit our model to the only one bremsstrahlung because the task to separate the two components is outside of the scope of the present paper. Such approach may overestimate somehow the plasma temperature but we are not interested in temperature (more detailed spatially resolved analysis of the X-ray spectra is performed e.g. in \cite{2024ApJ...962..159U}).} The {continuum is extrapolated to the lower energies and} used {to fit the bremsstrahlung around} the two lines. {Each} line is modelled as Gaussian (model \textit{gauss1d}) over the continuum {(this is why we do not need to know the temperature of the plasma)}. The \textit{gauss1d} model is characterized by three parameters: full width at the half maximum (\textit{fwhm}), the position of the center (\textit{pos}), and the amplitude (maximum peak) of the Gaussian (\textit{ampl}).  Examples of the fits are shown on Fig.~\ref{tycho3d:fig-linefits}. The values of \textit{rstat} {(that is the goodness of fit value $\chi^2$ divided by the number of degrees of fredom)} for most of the cells are below 10, for some cells they are between 10 and 20. If \textit{rstat} was above 20 (like shown on Fig.~\ref{tycho3d:fig-linefits} \textit{bottom}), the fit was corrected manually by narrowing the ranges for parameters and/or photon energy range for a particular fit. 

\begin{figure*}
  \centering 
  \includegraphics[trim=20 5 20 2, clip,height=6truecm]{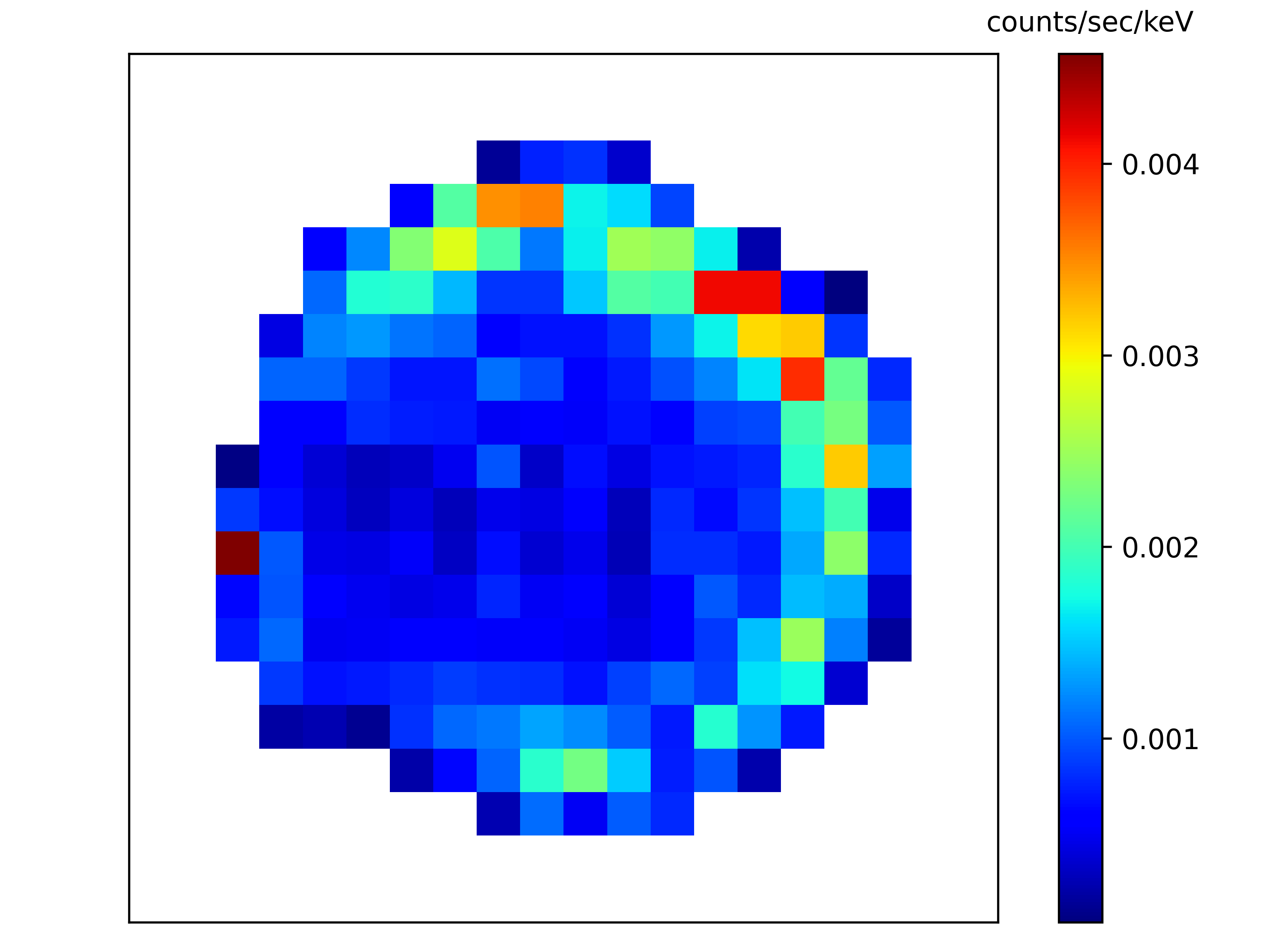} 
  \hspace{0.8cm}
  \includegraphics[trim=0 0 100 0, clip,height=6truecm]{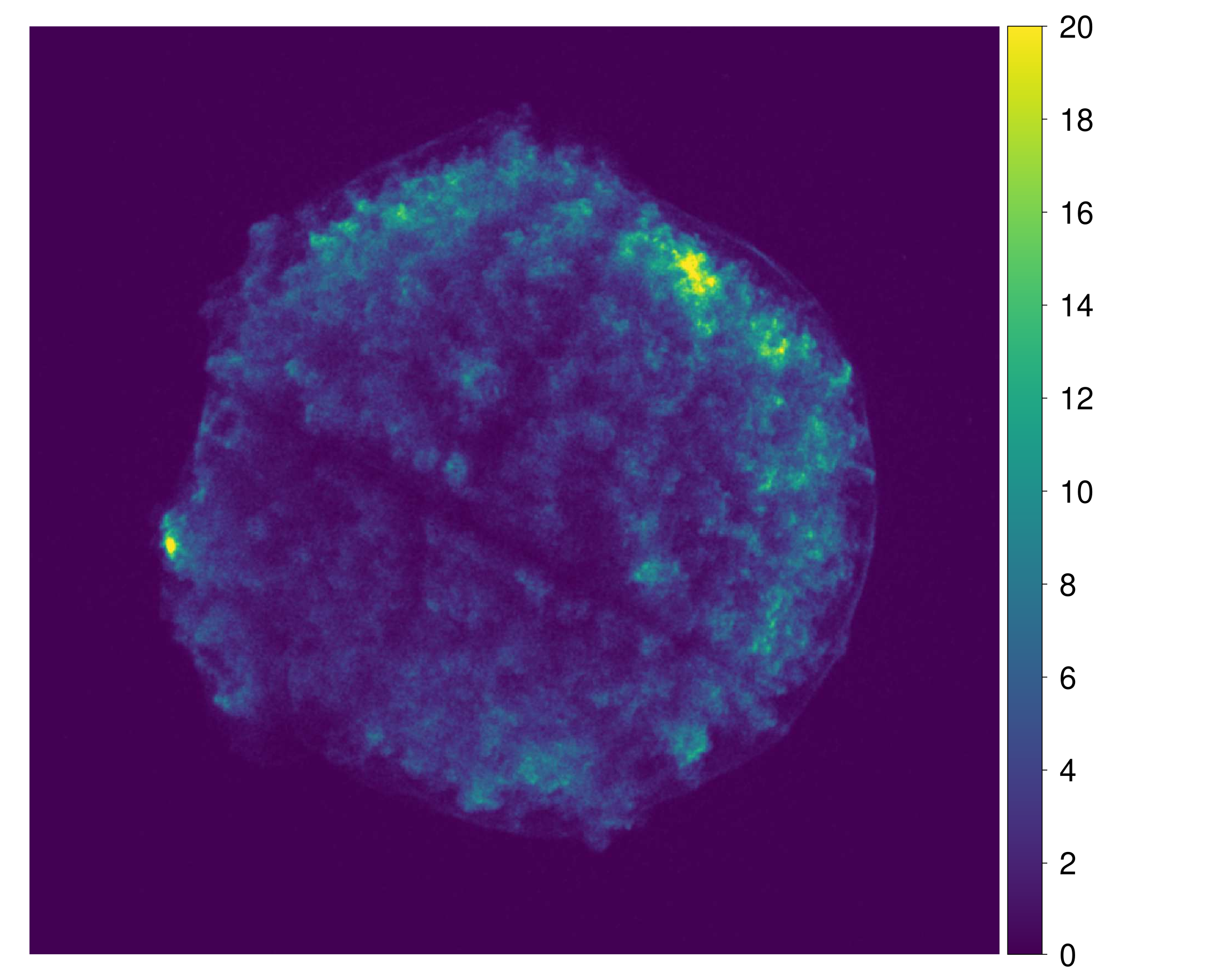}
  \caption{
    The map of the parameter \textit{ampl} from the model \textit{gauss1d} (\textit{left}) for the S line in comparison to the image of Tycho's SNR in the photon energy range $2.3-2.6\un{keV}$ (\textit{right}).
  }
  \label{tycho3d:fig-posvsimage}
\end{figure*}

Actually, we are interested in the map of the \textit{pos} parameter which equals to the central energy of the line $\varepsilon$. The two other (\textit{fwhm} and \textit{ampl}) are used to check the consistency of the derived results. In particular, \textit{fwhm} is largest ($\approx 0.10\un{keV}$ for Si and $\approx 0.18\un{keV}$ for S) for cells in the central part of SNR and decreases ($\sim 0.05\un{keV}$ for both lines) toward the edges, as expected because the line thickness reflects the level of randomness of velocities along the line of sight inside the object. As to another check, Fig.~\ref{tycho3d:fig-posvsimage} compares the distribution of the line amplitudes for each cell with the image of Tycho's SNR in photons with the relevant energies on the example of the S line. The good correspondence confirms that our approach is reliable.

\begin{table}
    \centering
    \caption{List of Chandra observations used for the dopplerography of Tycho's SNR}
    \begin{tabular}{ccc}
    \hline
    obsID  & start date & exposure time, ks \\
    \hline
    10093  & 2009-04-13 & 118.35 \\
    10094  & 2009-04-18 & 89.97  \\
    10095  & 2009-04-23 & 173.37  \\
    10096  & 2009-04-27 & 105.72  \\
    10097  & 2009-04-11 & 107.43  \\
    10902  & 2009-04-15 & 39.53  \\
    10903  & 2009-04-17 & 23.92  \\
    10904  & 2009-04-13 & 34.7  \\
    10906  & 2009-05-03 & 41.12  \\
    \hline
    \end{tabular}
    \label{tycho3d:table-observ}
\end{table}
\begin{figure*}
  \centering 
  \includegraphics[trim=20 5 20 2, clip,height=5.8truecm]{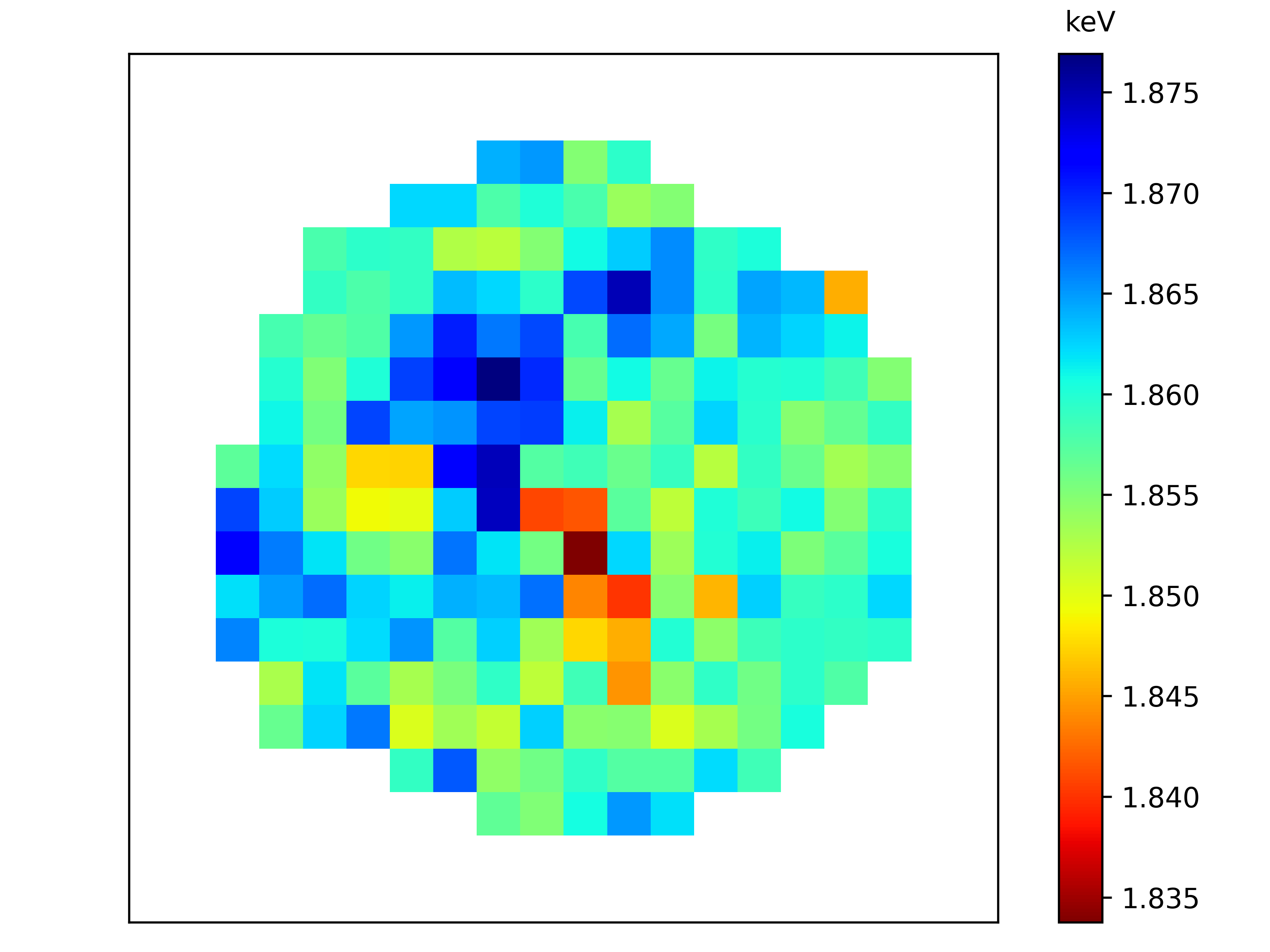}
  \includegraphics[trim=20 5 20 2, clip,height=5.8truecm]{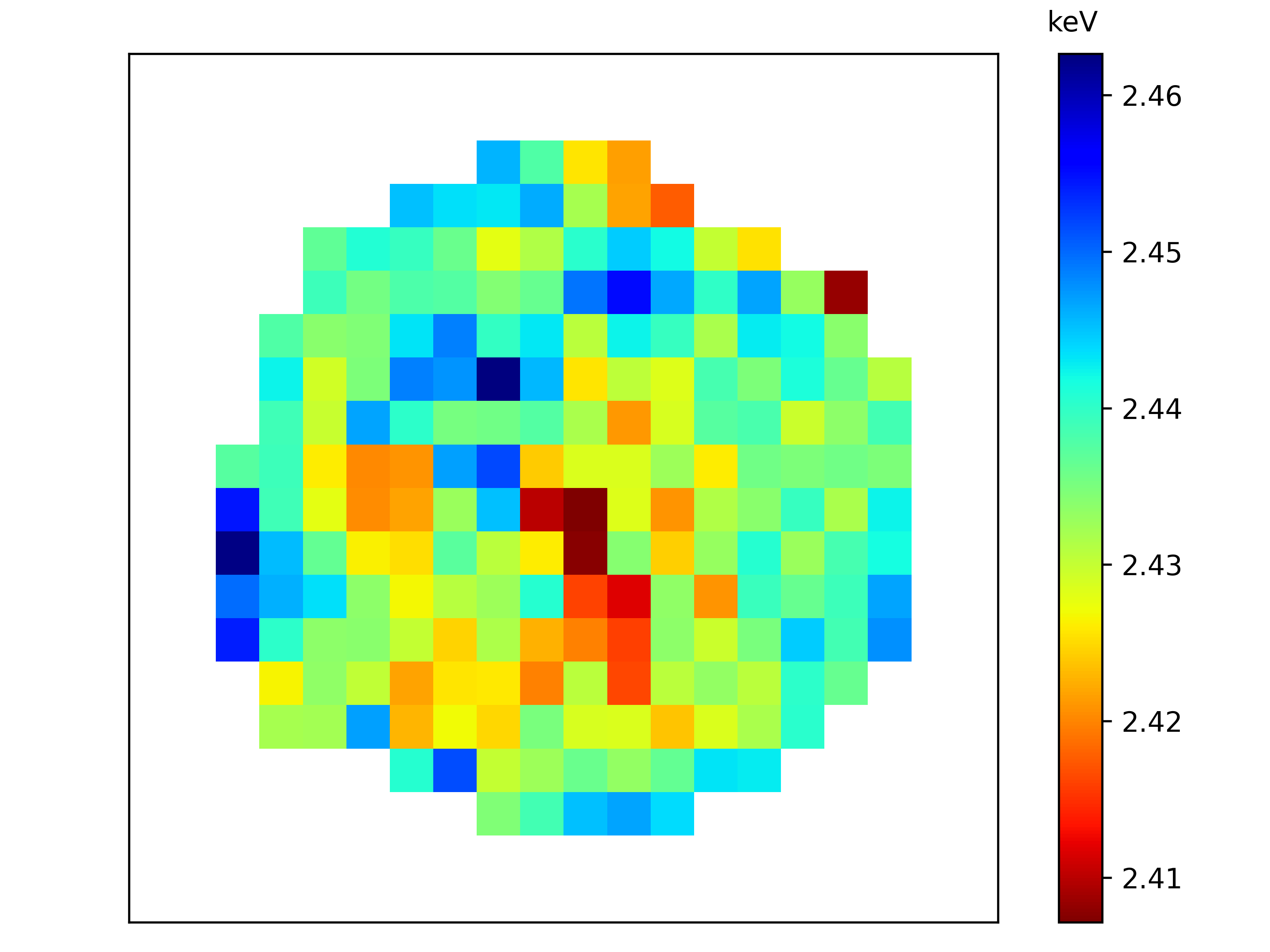}
  \caption{
    The map of the parameter \textit{pos} for the Si line (\textit{left}) and for the S line (\textit{right}). 
  }
  \label{tycho3d:pos_maps}
\end{figure*}
\begin{figure*}
  \centering 
  \includegraphics[trim=20 5 20 2, clip,height=5.8truecm]{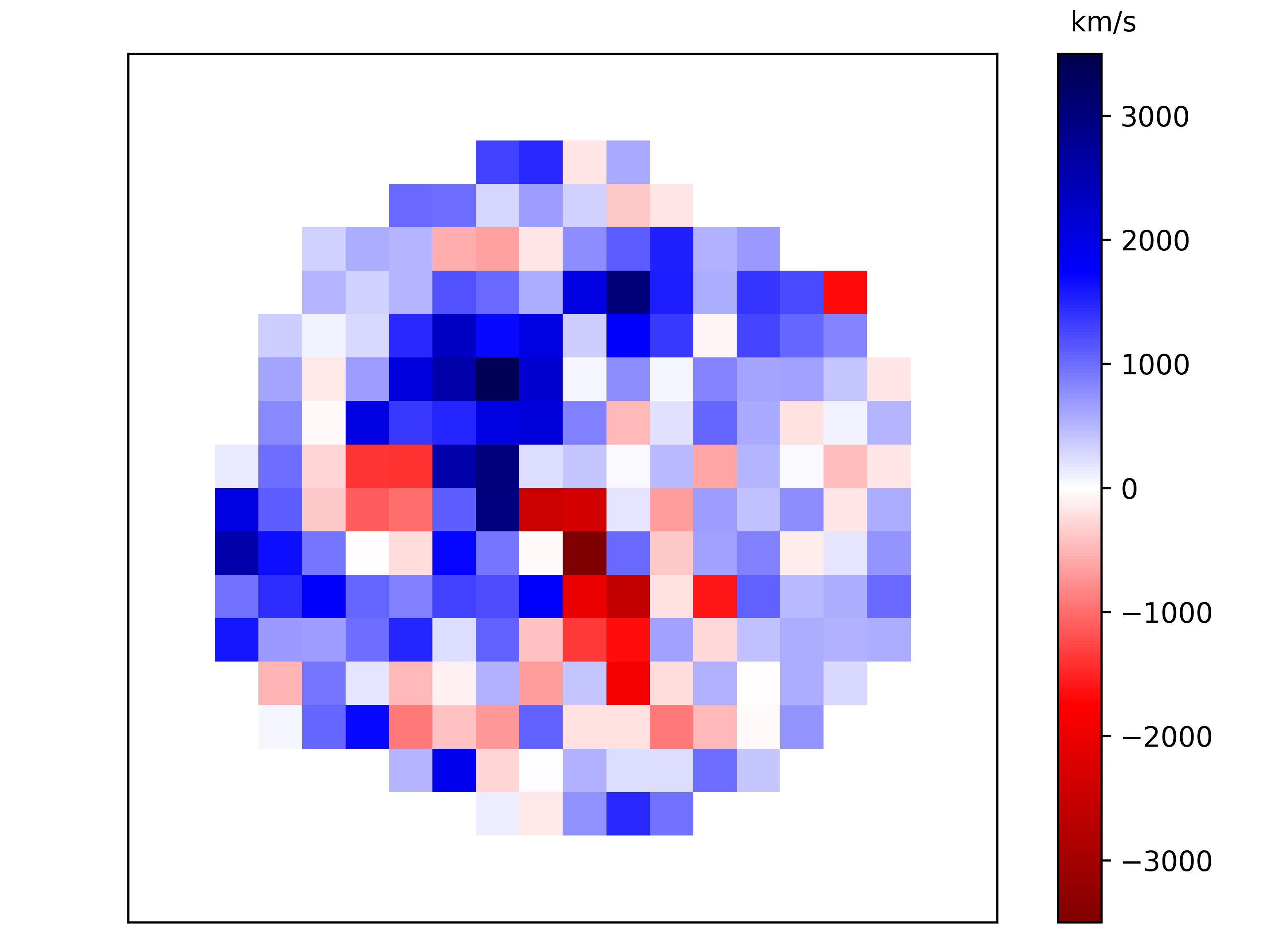}\ 
  \includegraphics[trim=20 5 20 2, clip,height=5.8truecm]{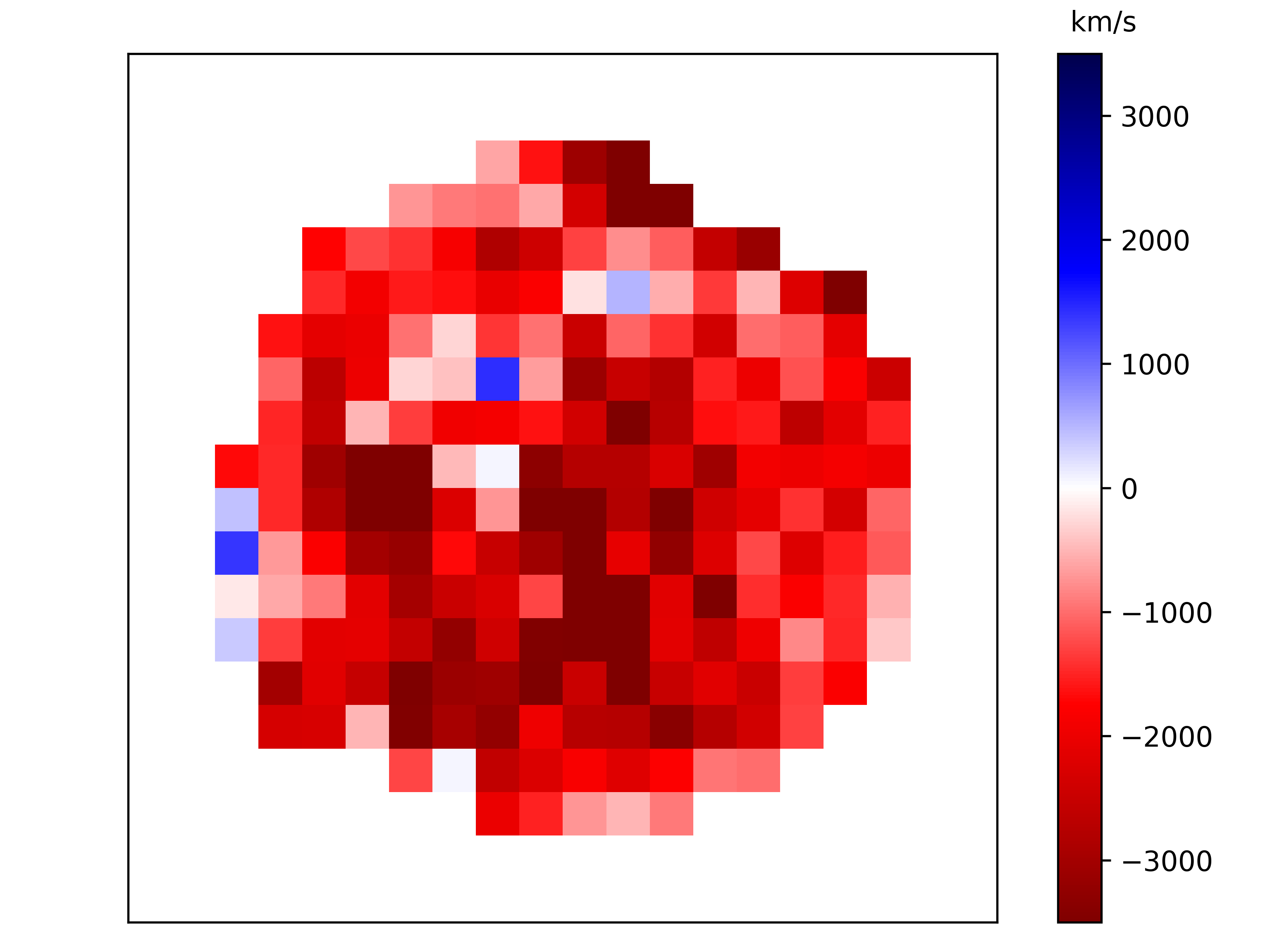}
  \caption{
    The map of the line-of-sight velocity for the Si line (\textit{left}) and for the S line (\textit{right}). The positive values of the velocity on the color scale correspond to the motion toward the observer. 
  }
  \label{tycho3d:v_maps}
\end{figure*}

Finally, Fig.~\ref{tycho3d:pos_maps} demonstrates the maps of the central energy $\varepsilon$ for the lines Si and S. 
Our map for the Si line is in good agreement with Fig.~4 in \cite{2017ApJ...840..112S} where the mean photon energy map in the Si–He$\alpha$ band (1.6-2.1 keV) is presented as well as with Fig.~4 in \cite{2023A&A...680A..80G} where the map of the peak energy for Si line is reported. Such agreement testifies the robustness of our approach which differs from the two cited above. Namely, we performed a detailed fit of the line shape for each cell in our grid.

The map of the central energy $\varepsilon$ may be converted to the map of velocities along the line of sight $v\rs{\|}$. Namely, the Doppler effect for the observer at rest yields 
\begin{equation}
    v\rs{\|}=c\frac{\varepsilon-\varepsilon\rs{o}}{\varepsilon\rs{o}}
\end{equation}
where $c$ is the speed of light, $\varepsilon\rs{o}$ is the `laboratory' energy, i.e. the central energy of the line for the source at rest. The velocity $v\rs{\|}>0$ (`blue shift') if the material moves toward the observer.
The transition from $\epsilon$ to $v\rs{\|}$ is affected by the value of the `laboratory' velocity. We take the `laboratory' energy as $1.856\un{keV}$ for the Si and $2.450\un{keV}$ for the S line (Appendix \ref{Tycho3D:app2}). 

Fig.~\ref{tycho3d:v_maps} shows the maps for the line-of-sight velocity for the ejecta material, which is rich in Si and S ions. The image for silicon (on the left) correlates very well with analogous image shown on Fig.~4 in \cite{2023A&A...680A..80G} {and Fig.~5 (left middle plot) in \cite{2024ApJ...962..159U}} which were derived in different approaches. 
There are regions with Si which run to and out of the observer. There is some asymmetry between the north-east (there are more Si material expanding toward us) and the south-west half of SNR. Instead, most of the material consisting of S seems to move out of the observer. This could be a sign of asymmetry of SN explosion. 

\begin{table}
  \centering 
  \caption{The parameter \textit{pos} and the LoS average velocity $v$ for the Si and S material in four regions of the SNR. SNR is divided into quadrants by a Cartesian coordinate system centered on the geometrical center with the vertical axis directed to the north. Therefore, the quadrant q1 is the upper right one, etc. 
  }
  	\begin{tabular}{l|cc|cc} 
  \hline
  region &\multicolumn{2}{c}{pos, keV}& \multicolumn{2}{|c}{$v$, km/s}\\
   &Si &S& Si & S \\
  \hline
  q1 &1.857	&2.449 &210	&-110\\
  q2 &1.861	&2.449 &743	&-73\\
  q3 &1.861	&2.446 &840	&-477\\
  q4 &1.856	&2.441 &32	&-1077\\
  SNR &1.858	&2.448 &420	&-269\\
  \hline
  lab &1.856	&2.450 &0 &0\\
  \hline
	\end{tabular}
  \label{tycho3d:epsvpie}
\end{table}

For the sake of comparison, we have performed the same kind of analysis for the Tycho's SNR as a whole and found that the value of the \textit{pos} parameter is $1.858\un{keV}$ for Si and $2.448\un{keV}$ for S line. They are quite close to the laboratory energies, therefore, the shocked Si and S material as a whole is not moving along the line of sight (LoS). The parameter \textit{fwhm} for the total emission from Tycho's SNR is $0.065\un{keV}$ for Si and $0.081\un{keV}$ for S. 

At first glance, it could be strange that e.g., most of the cells are red- in S or blue-shifted in Si line (Fig.~\ref{tycho3d:v_maps}) while the overall velocity of SNR is close to the `laboratory' energy, for both lines. However, the velocity for a cell is a sort of an average along the line of sight; it accounts for motions in the front and the rear half-spheres. The blue and red shift itself does not mean that the whole material moves in one direction. It means that there is a slight advance in motion in one direction over the whole LoS inside SNR. In addition, the plasma elements which are more bright contribute more to the LoS velocity, i.e. they dominate the `color' of a cell. The variation of abundance may also affect the variation of average velocities between the quadrants. for example, the equivalent width map for the Si emission lines in the 1.65-2.05 keV band shows the variation of Si abundance within the factor of two over the projection of Tycho's SNR (see Fig.~5 in \cite{2024ApJ...972...63P}).

We have repeated the same analysis for the four quadrants of the remnant of the Tycho supernova (Table~\ref{tycho3d:epsvpie}). The differences in the average LoS velocities between different species in the same region is a strong sign of asymmetries in their 3-D spatial distributions that could be due to asymmetry in the explosion as well as mixing of layers in the progenitor star prior to the explosion. 

\begin{figure}
  \centering 
  \includegraphics[width=7.5truecm]{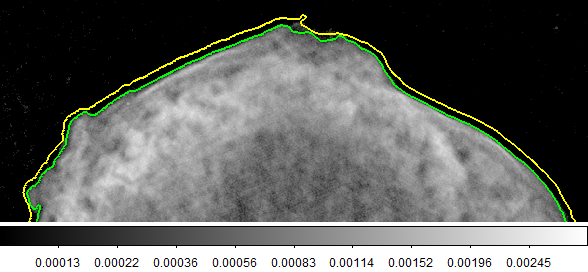}
  \caption{
    North of Tycho's SNR. Gray-scale shows the square root of the radio brightness at $\sim 1.4\un{GHz}$ for the epoch 1994. Green and yellow contours correspond to the years 1994 and 2013 respectively. They trace the edge of SNR at the brightness $3\E{-4}\un{Jy}$. 
  }
  \label{tycho3d:contours}
\end{figure}

\begin{table}
  \centering 
  \caption{Shock speed $V$ at the periphery of Tycho's SNR in km/s estimated with the procedure described in Sect.~\ref{tycho3d:vyz-density} in comparison with determined in \cite{2013ApJ...770..129W} for the same assumption about the distance to the SNR. 
  Our statistical uncertainties correspond to the $1\sigma$ level.
  }
  	\begin{tabular}{lcc} 
  \hline
  region & present study& \cite{2013ApJ...770..129W}\\
  \hline
  q1 &$3500\pm460$	&$3300\pm330$\\
  q2 &$3700\pm830$	&$3700\pm340$\\
  q3 &$3600\pm520$	&$3300\pm400$\\
  q4 &$3900\pm470$	&$3900\pm520$\\
  SNR &$3700\pm600$	&$3300\pm390$\\
  \hline
	\end{tabular}
  \label{tycho3d:tableVave}
\end{table}


\section{3D velocity fields for Si and S}
\label{tycho3Da:velocities}

In this section, we aim to reconstruct the 3D structure of the flow velocity inside Tycho's SNR.  

\subsection{Expansion of SNR in the plane of the sky}
\label{tycho3d:vyz-density}

The shock speed $V$ may be estimated by analyzing the proper motion of the SNR edge. 
It varies with azimuth in Tycho's SNR. Authors in \cite{2013ApJ...770..129W} listed the values of $V$ for a number of azimuth angles. Their estimates are based on a kinematic studies of Tycho's SNR in radio band \cite{1997ApJ...491..816R} and in X-rays \cite{2010ApJ...709.1387K}. We need estimates for the shock speed and also for other angles. Therefore, instead of interpolation of values from \cite{2013ApJ...770..129W}, we follow a simple procedure. Namely, we use the VLA radio images of the remnant at 1.4 GHz from the years 1994 and 2013 \cite{2016ApJ...823L..32W} and determine the velocity as $V=\Delta R/\Delta t$ where $\Delta R$ is the difference of radii at a given azimuth from 1994 to 2013 year and $\Delta t=19\un{yrs}$. In order to determine the radii, we draw the contour lines on both radio images at the same level of brightness, $3\E{-4}\un{Jy}$ that is $1/10$ of the maximum brightness on both images. The outer cut-off of the radial profile is rather sharp at the SNR edge; therefore, the shock position is reasonably tracked in this way. In order to lower the fluctuations on a pixel level, we smoothed the images a bit by setting \verb+smoothness+ in \verb+ds9+ to 10 pixels. The size of the images is $4096\times 4096$ pixels, therefore, the effect of such smoothing to the overall image is almost negligible. Fig.~\ref{tycho3d:contours} shows the northern region of SNR with the two contours used to determine the shock speed.
We place the center of the explosion at the coordinates determined in \cite{2015ApJ...809..183X} and measure the radii $R$ from this point. The distance to Tycho's SNR is taken as $d=2.3\un{kpc}$ \cite{2013ApJ...770..129W}. 

In this way, we have determined the shock speed over the whole edge of the SNR projection. It is close to the values given in \cite{2013ApJ...770..129W} for different azimuths. In particular, the average values are shown in Table~\ref{tycho3d:tableVave}. The values agree well within the errors.

\begin{figure}
  \centering 
  \includegraphics[width=7.5truecm]{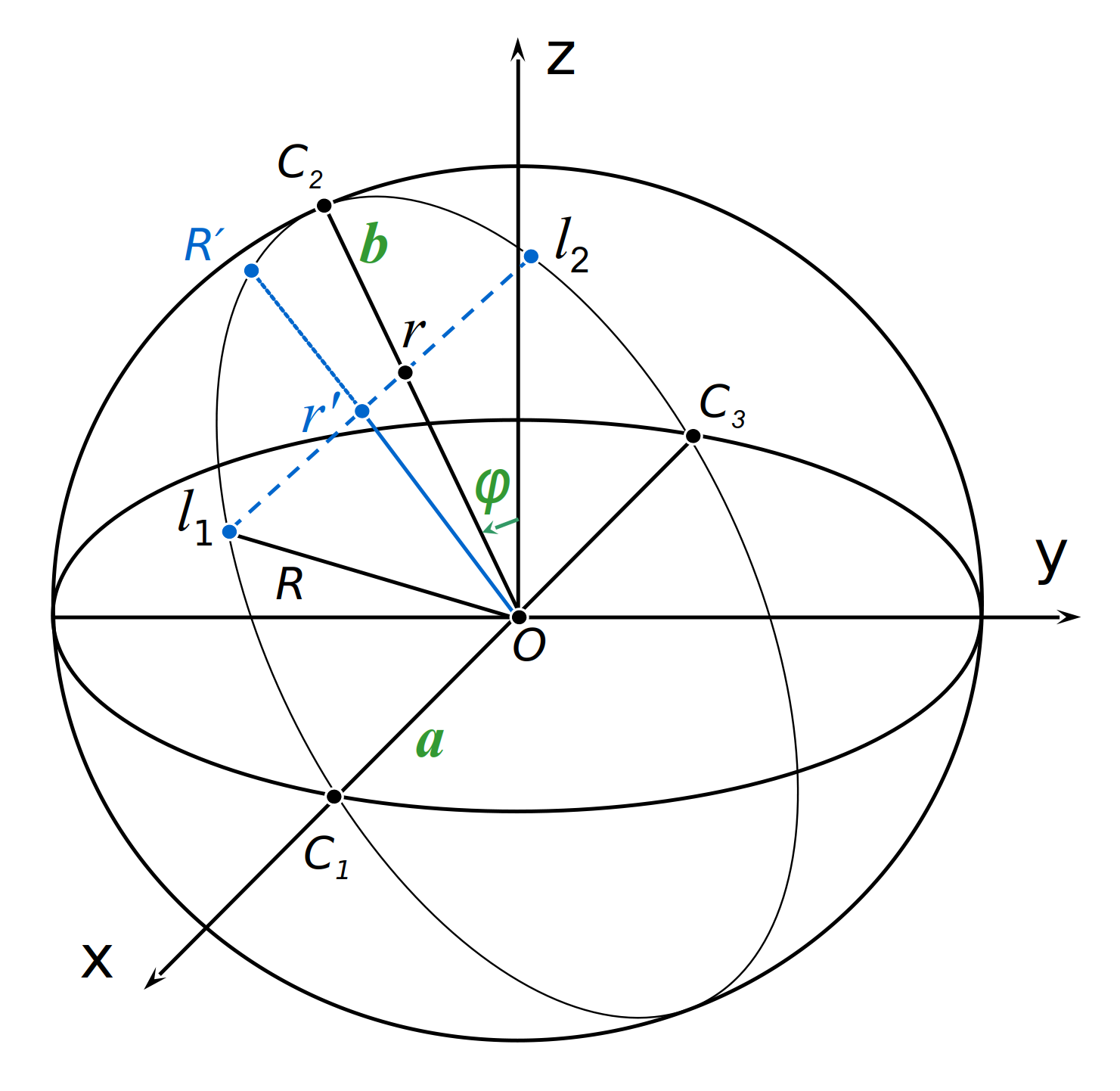}
  \caption{
    Geometry of the task. Axis $x$ is directed toward the observer, $z$ to the north, $\phi$ is an angle in the projection plane.
  }
  \label{tycho3Da:fig_sxema}
\end{figure}
\begin{figure*}
  \centering 
  \includegraphics[trim=30 34 26 10,clip,width=0.96\textwidth]{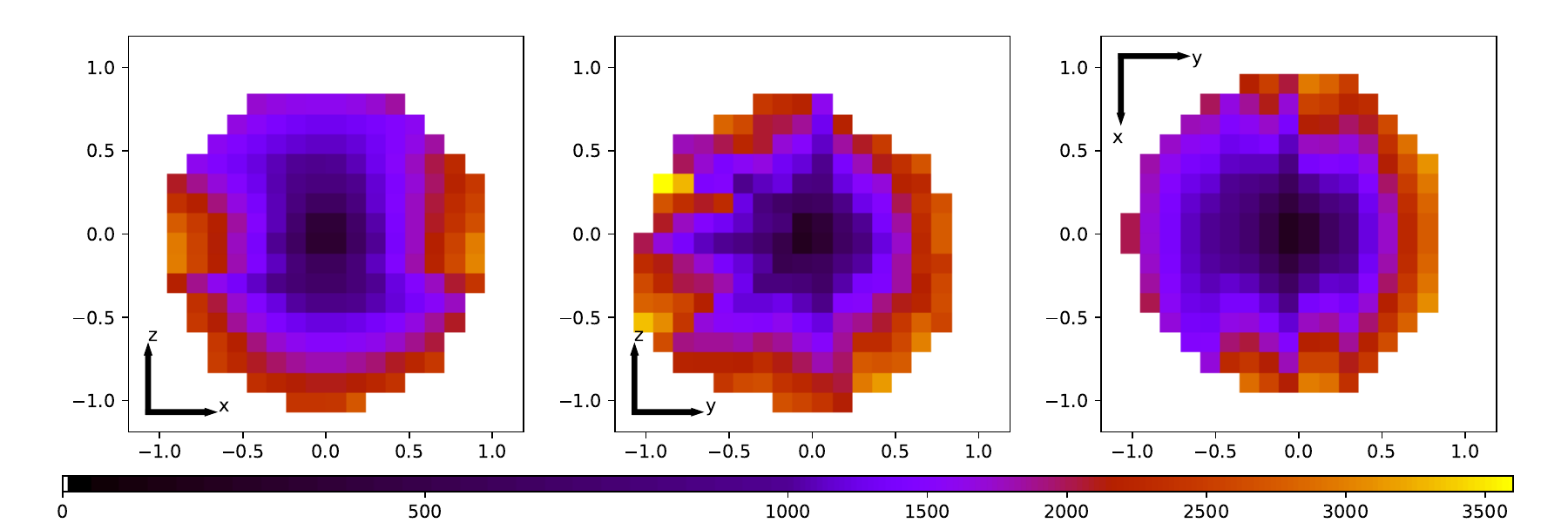}
  \includegraphics[trim=30 00 26 10,clip,width=0.96\textwidth]{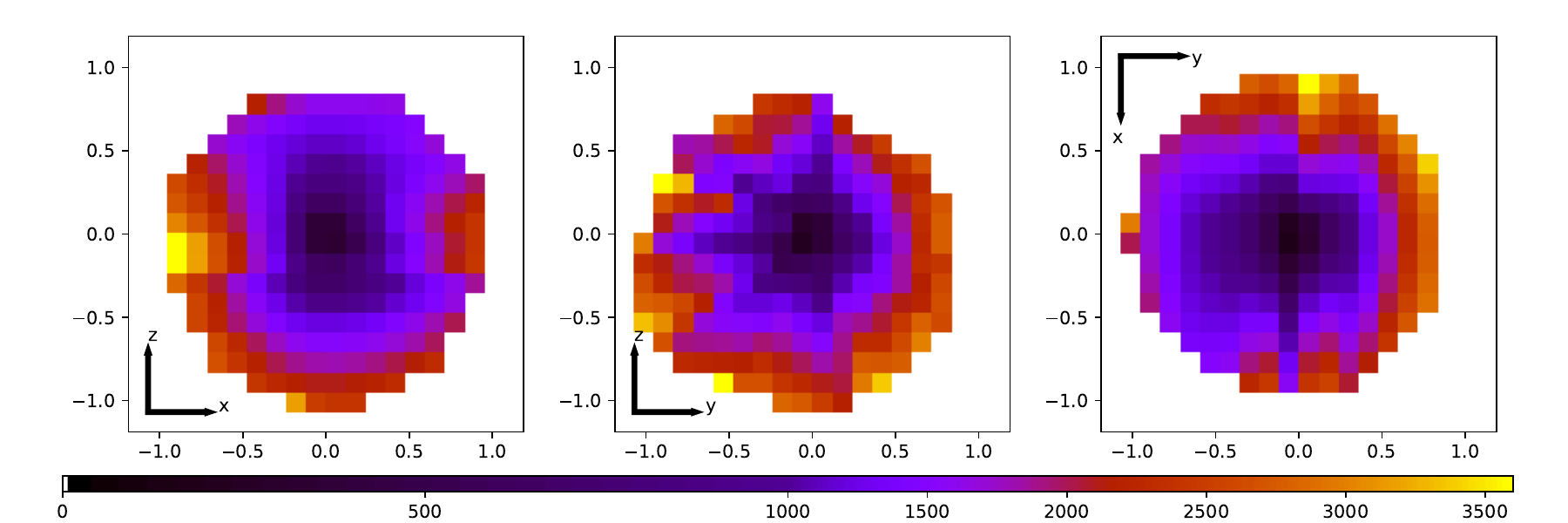}
  \caption{
    Cross-sections of the velocity data cubes by three planes passing through the explosion center for Silicon (top) and Sulfur (bottom). The line of sight is parallel to the $x$-axis. 
  }
  \label{tycho3Da:fig_crosssectS}
\end{figure*}
\begin{figure*}
  \centering 
  \includegraphics[width=0.9\textwidth]{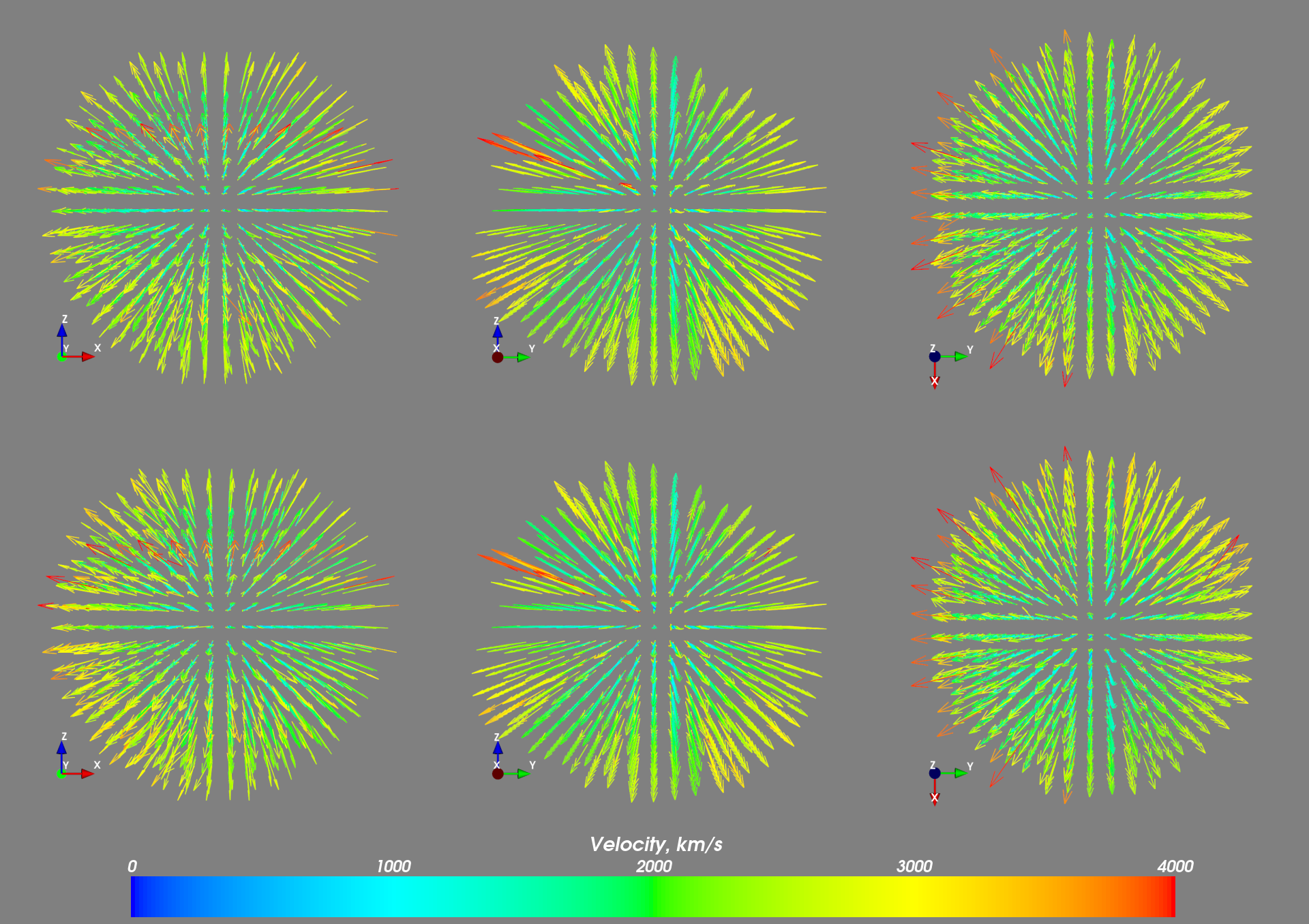}  
  \caption{
    The 3D velocity fields for Si (top panels) and S (bottom panels). 
  }
  \label{tycho3Da:fig_fieldsSSi}
\end{figure*}

\subsection{3D vector field}

Now, we have the observational information to reconstruct the 3D velocity field in Tycho's SNR. Namely, the component along the LoS from dopplerography and the component in the PoS from the proper motion. We will use these data together with the Abel inversion and some assumptions.  

Let us consider the Cartesian coordinate system with the center located in the explosion point, the axis $x$ is directed to the observer, $y$ and $z$ are in the plane of the sky (Fig.~\ref{tycho3Da:fig_sxema}). 
The radii $OC_2$ of SNR projection onto the plane of the sky are different for different azimuth angles $\varphi$. 
The size of SNR along the line of sight is unknown. Therefore, we take that the radius $OC_1$ equals to the average radius in $yz$ plane. 
A half-ellipse $C_1C_2C_3$ has the axis $a$ along the line of sight and another axis $b$ perpendicular to it (Fig.~\ref{tycho3Da:fig_sxema}). An SNR radius $R'$ in the $C_1C_2C_3$ plane varies from $a$ to $b$ as  
\begin{equation}
 R' = \sqrt{x^2 \left( 1 - {b^2}/{a^2} \right) + b^2}.
\end{equation}

To start with, we build the 3D data cube for the flow velocity distribution, which accounts for the asymmetry in the $yz$ plane but is symmetric relative to the plane of the sky. Namely, we assume at this step that i) the shock speed along each radius $R'$ in the $C_1C_2C_3$ cross-section is the same and equals to the speed measured for the direction $OC_2$ in the plane of the sky and ii) the expansion is radial. 

The flow speed in young SNR is commonly taken as proportional to the distance from the center \cite{1950RSPSA.201..159T,1982ApJ...259..302C}. We adopt the same approximation, namely, the radial profile of the flow speed in a point $r'$ along a radius is $v=v\rs{s}r'/R'$ where $v\rs{s}=3V/4$ is the flow speed immediately downstream, $V$ is the shock speed for a given azimuth.  
The components of the velocity $\mathbf{v}$ in a point $(x,y,z)$ inside SNR are $v\rs{x}=vx/r'$, $v\rs{y}=vy/r'$, $v\rs{z}=vz/r'$. For example, we have for the LoS component 
\begin{equation}
 v_x = \frac{x}{r'} v_s \frac{r'}{R'(x,\varphi)} = \frac{3xV(\varphi)}{4R'(x,\varphi)}
 \label{tycho3d:vlos}
\end{equation}

As the second step, we introduce the LoS asymmetry into this plane-symmetric 3D velocity field by using the results of the dopplerography from Sect.~\ref{tycho3Da:sect-doppl}. 
Doppler effect is sensitive to the $v\rs{x}$ component of $v$.
The velocity $V\rs{D}$ in each cell on Fig.~\ref{tycho3d:v_maps} is a residual of the sum of the $v\rs{x}$ components along the line of sight. The components $v\rs{x}>0$ in the nearest half-space (i.e. for $x>0$) and  $v\rs{x}<0$ for $x<0$. 
$V\rs{D}=0$ in a model where the velocity distribution is symmetric relative to the $yz$ plane. 
Therefore, we change the local values of $v\rs{x}$ along a given line of sight to $v\rs{x}'=\alpha v\rs{x}$ in the nearer (farther) half of SNR and to $v\rs{x}'= v\rs{x}/\alpha$ in the farther (nearer) half for $V\rs{D}>0$ ($V\rs{D}<0$), i.e.
\begin{equation} \label{V_D_1}
    V\rs{D} = \sum_{x>0} v'_x - \sum_{x<0} |v'_x| = \sum_{x>0} a v_x - \sum_{x<0} \frac{1}{a} |v_x| = \left( a - \frac{1}{a} \right) S
\end{equation}
where we denoted the sum of $v\rs{x}$ along the LoS in one half-space  
\begin{equation} 
    S = \sum_{x>0} v_x = \sum_{x<0} |v_x| .
\end{equation}
The value of $\alpha$ is given by a solution of a quadratic equation
\begin{equation} \label{a_1}
    \alpha = \frac{|V_D|}{2S} + \sqrt{\frac{V_D^2}{4S^2} + 1}.
\end{equation}
In this way, we derive the 3D velocity fields for Si and S which 
account for the asymmetries in the projection plane $yz$ and along the LoS, which is parallel to the $x$-axis. 

Our final results are presented on Figs.~\ref{tycho3Da:fig_crosssectS} and \ref{tycho3Da:fig_fieldsSSi}. Fig.~\ref{tycho3Da:fig_crosssectS} shows the cross-sections of the velocity data cube for the ejecta material with {Silicon-rich and Sulfur-rich plasma (a few times higher of the Solar abundance)}. Note the asymmetries along the $x$-axis, which reflect the Doppler shifts of the emission line, e.g., by summing up velocities along the $x$-axis, S material comes out to be mostly red-shifted as observed. 
Fig.~\ref{tycho3Da:fig_fieldsSSi} shows 3D vector fields for the velocities of the material consisting of Si and S. 
Note that the flow velocity vectors are not radial everywhere, though the radial component dominates. Another important conclusion is that, despite the strong `red-shifted' image for Sulfur on Fig.~\ref{tycho3d:v_maps}, the differences between absolute values of the flow velocities in the front and the rear half-spheres are not dramatically large but just on the level of $20-30\%$ (Fig.~\ref{tycho3Da:fig_fieldsSSi}). Thus, the asymmetry in the supernova explosion should not be excessive to provide the observed distribution of the Doppler shifts.

{One can notice an interesting feature on Fig.~\ref{tycho3Da:fig_fieldsSSi}. Namely, there are vectors with the absolute values of the plasma velocity $>3500\un{km/s}$. The average shock velocity (Table~\ref{tycho3d:tableVave}) in quadrants is $3500\div 3900\un{km/s}$. The plasma speed should be at most $3/4$ of this value, i.e. up to $2900\un{km/s}$. 
However, the values for the plasma speed $>3500\un{km/s}$ could be present naturally because (i)  $2900\un{km/s}$ is an average number (so, there are also higher values), (ii) uncertainty in the plasma speed determination due to the Doppler effect is of order $500\un{km/s}$ \cite{2023A&A...680A..80G}, (iii) both methods (from Doppler shifts and proper motion) determine the components of the velocity, their vector sum is higher than each of the component. In addition, in some regions the shock may encounter the dense wall of stellar wind bubble as proposed in \cite{2024ApJ...962..159U}. In this case, the shock should decelerate in the direction of its normal but the following plasma may be forsed to move rapidly in a perpendicular direction.}

\section{Conclusions}

We have reconstructed the 3D vector field of the velocity {of Si- and S-rich plasma} in the remnant of Tycho supernova. The components of velocity vectors along the LoS were determined from the maps of Doppler shifts of the Si and S lines in the X-ray spectrum. In order to produce these maps, we performed a spatially resolved spectral analysis of the data of X-ray observations by the Chandra Observatory. Other components of the velocity vectors, in the PoS, were determined from the proper motion of the outer layers of the remnant. We used the radio observations of Tycho's SNR derived with VLA \cite{2016ApJ...823L..32W}. In order to restore the absolute values of the components in the SNR interior, we adopted some theoretical considerations. Firstly, {we used a rather simple procedure based on the} Abel inversion {but for elliptic cross-sections}. Secondly, we used the assumption of the homologous expansion of the ejecta in young SNRs where $v\propto r$, which is based on analytical solutions {for either the ejecta-dominated stage \cite{1982ApJ...259..302C} or Sedov-Taylor one  \cite{1950RSPSA.201..159T}} and is widely used in the literature {for reconstruction of velocities from Doppler shifts in SNRs}. 

Generally, the 3D velocity fields for Si and S look similar in Tycho's SNR. However, {we found for the first time that} there are differences in the 3D spatial distributions between the Si-reach and the S-reach ejecta components. Indeed, the sulfur material is more red-shifted while the distribution of Si is more isotropic (Figs.~\ref{tycho3d:v_maps}, \ref{tycho3Da:fig_crosssectS}, \ref{tycho3Da:fig_fieldsSSi}).

The 3D asymmetry in the supernova explosion is moderate to account for the observed distribution of the Doppler shifts over the SNR projection. In fact, the differences in the absolute values of the Doppler velocities are up to a few thousand km/s. Differences in the plasma speed are about $20-30\%$ on the opposite sides of SNR as a 3D object. 

Such findings reveal a possible level of anisotropy in the supernova explosion and provide an evidence about deviation of the internal structure of a star from a spherically-symmetric layered one. This may be caused by `sloshing' and/or mixing inside a progenitor prior to a supernova event. 
Our results may be used as input for numerical simulations of Tycho's SNR evolution.

\section{Acknowledgements}
We acknowledge Bohdan Melekh for discussions on the line emission diagnostics, Marco Miceli for advice on the X-ray data analysis, Laura Chomiuk for the radio image of Tycho's SNR, Adam Foster for the help in the usage of AtomDB.  
M.P. acknowledges the hospitality and financial support of Astronomical Observatory in Jagiellonian University for training on X-ray data analysis during International Summer Student Internships as well as the support from INAF 2023 RS4 minigrant. We thank the Armed Forces of Ukraine for providing security to perform this work. 
This project has received funding through the MSCA4Ukraine project, which is funded by the European Union. Views and opinions expressed are however those of the author(s) only and do not necessarily reflect those of the European Union. Neither the European Union nor the MSCA4Ukraine Consortium as a whole nor any individual member institutions of the MSCA4Ukraine Consortium can be held responsible for them.


\begin{thebibliography}{10}
	
	\bibitem{2005NatPh...1..147W}
	S.~{Woosley}, T.~{Janka}, 
	{\em Nature Physics} \textbf{1}, 147 (2005)
	
	\bibitem{2008ApJ...681.1448J}
	I.~{Jordan} {\it et al.}, {\apj} \textbf{681}, 1448 (2008)
	
	\bibitem{2015A&A...577A..48W}
	A.~{Wongwathanarat}, E.~{M{\"u}ller}, H.~T. {Janka}, { \aap} \textbf{577}, A48 (2015)
	
	\bibitem{2023MNRAS.518.1557R}
	M.~{Reichert} {\it et al.},  { \mnras} \textbf{518}, 1557 (2023)
	
	\bibitem{2017hsn..book..117D}
	A.~{Decourchelle}, ``{Supernova of 1572, Tycho's Supernova}'' in {\it Handbook
		of Supernovae} (Springer, 2017), p.117
	
	\bibitem{2017ApJ...842...28W}
	B.~J. {Williams} {\it et al.},  { \apj}
	\textbf{842}, 28 (2017)
	
	\bibitem{2017ApJ...840..112S}
	T.~{Sato}, J.~P. {Hughes},  { \apj} \textbf{840}, 112 (2017)
	
	\bibitem{2023A&A...680A..80G}
	L.~{Godinaud}, F.~{Acero}, A.~{Decourchelle}, J.~{Ballet},  { \aap} \textbf{680}, A80 (2023)
	
	\bibitem{2024arXiv240417296G}
	L.~{Godinaud}, F.~{Acero}, A.~{Decourchelle}, J.~{Ballet},  { arXiv e-prints}, 2404.17296 (2024)
	
	\bibitem{2024ApJ...962..159U}
	H.~{Uchida} {\it et al.},
	 { \apj} \textbf{962}, 159 (2024)
	
	\bibitem{1997ApJ...491..816R}
	E.~M. {Reynoso} {\it et al.},  { \apj} \textbf{491}, 816 (1997)
	
	\bibitem{2010ApJ...709.1387K}
	S.~{Katsuda} {\it et al.},  { \apj} \textbf{709}, 1387 (2010)
	
	\bibitem{2013ApJ...770..129W}
	B.~J. {Williams} {\it et al.},  {
		\apj} \textbf{770}, 129 (2013)
	
	\bibitem{2021ApJ...906L...3T}
	T.~{Tanaka} {\it et al.},  { \apjl} \textbf{906}, L3 (2021)
	
	\bibitem{2016ApJ...823L..32W}
	B.~J. {Williams} {\it et al.},  { \apjl} \textbf{823}, L32 (2016)
	
	\bibitem{2024ApJ...972...63P}
	O.~{Petruk} {\it et al.},  { \apj} \textbf{972}, 63 (2024)
	
	\bibitem{2015ApJ...809..183X}
	Z.~{Xue}, B.~E. {Schaefer},  { \apj} \textbf{809}, 183 (2015)
	
	\bibitem{1950RSPSA.201..159T}
	G.~{Taylor}, {Proc. Royal Soc. London
		Ser. A} \textbf{201}, 159 (1950)
	
	\bibitem{1982ApJ...259..302C}
	R.~A. {Chevalier}, 
	{ \apj} \textbf{259}, 302 (1982)
	
	\bibitem{2012ApJ...756..128F}
	A.~R. {Foster}, L.~{Ji}, R.~K. {Smith}, N.~S. {Brickhouse},  {\apj} \textbf{756}, 128 (2012)
	
\end{thebibliography}


\section{Appendix}

\subsection{Rest energies for the lines}
\label{Tycho3D:app2}

In order to translate the observed shifts in the energies of Si and S lines into the line-of-sight velocities, we need to know the reference (`laboratory' or `rest') energy of the lines. We used the AtomDB database 
\cite{2012ApJ...756..128F} to derive its value. It follows from analysis with AtomDB that most photons around 1.86 and 2.45 keV arise from transitions $7\rightarrow 1$, $6\rightarrow 1$, $5\rightarrow 1$, $2\rightarrow 1$ in both Si~\textsc{XIII} and S~\textsc{XV} ions. Other ions contribute on the level of a few per cent around these energies and may be neglected. 

\begin{figure}
  \centering 
  \includegraphics[trim=25 10 55 25,clip,width=7.5truecm]{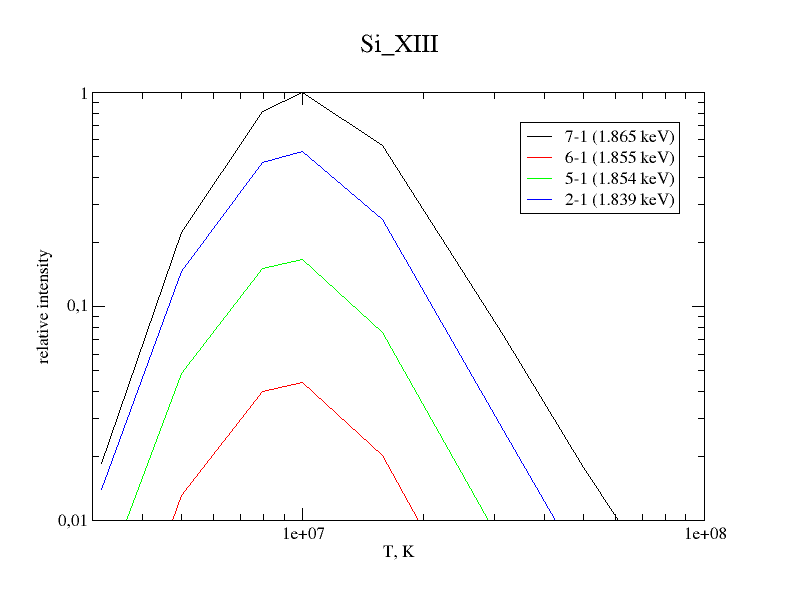}
  \includegraphics[trim=25 10 55 25,clip,width=7.5truecm]{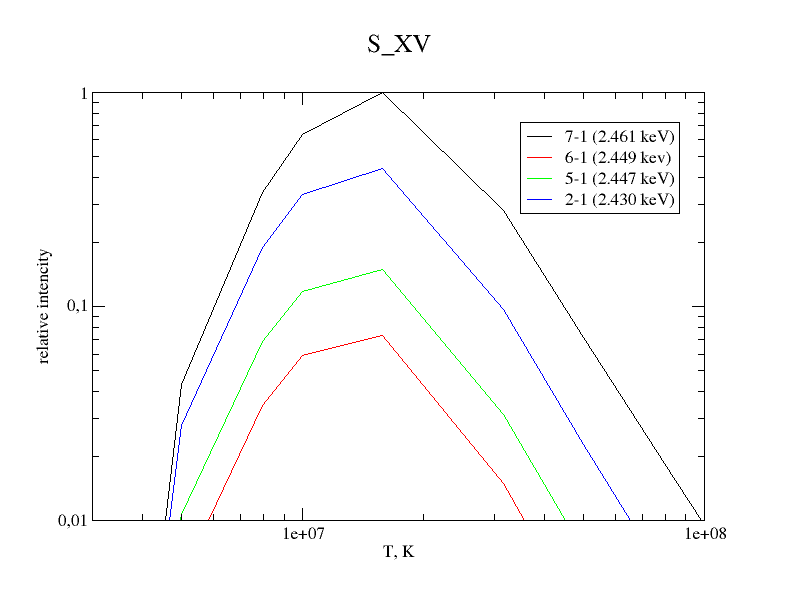}
  \caption{
    Intensity of emission for transitions (shown by colors) for Si~\textsc{XIII} (\textit{top}) and S~\textsc{XV} (\textit{bottom}) ions at different temperatures. Vertical axes show values normalized to the maximum value for the black lines.
  }
  \label{tycho3Da:fig_rint}
\end{figure}

Fig.~\ref{tycho3Da:fig_rint} shows that i) the two transitions, $7\rightarrow 1$, $2\rightarrow 1$, are dominant in both lines; ii) the {\it relative} intensities of the transitions vary slightly with $T$ in the temperature range where the lines are visible. We also see that if a relative intensity for a given two transitions is less than unity at some temperature, it does not become larger than unity for other $T$. Therefore, the observed effect of the line shift over the SNR surface should be due to a Doppler effect, not due to different conditions in various regions in SNR.

In order to determine the peak energy, we plot Gaussians representing the lines for these four transitions at their own energies and with the half-width, which restores the resolution of the observational data. The amplitudes of the Gaussians for each transition are taken with relative intensities at $T=10^{7}\un{K}$. We then add all four contributions and fit the sum by a Gaussian with the same half-width. The position of its maximum is taken as a `laboratory' energy of the line. 

The resulting energies are $1.856$ keV for Si~\textsc{XIII} and $2.450$ keV for S~\textsc{XV}. Though the relative intensities vary somehow with temperature, this `laboratory' energy is almost insensitive to it. For example, the energies are $1.855$ and $2.449$ keV for relative ratios taken at $T=5\E{6}\un{K}$, difference of $\sim 1\un{eV}$. 

We performed this analysis assuming the collisional ionization equilibrium. The peak energy could be affected somehow by the ionization state (represented by the ionization parameter $\tau=n\rs{e}t$) and response matrix of the ACIS instrument of the Chandra observatory. We checked this out (also considering the effect of the instrument). Generally, if $\tau$ changes on two orders of magnitude, the rest energy changes on $\sim 15$ eV that is less than $1\%$ (M.Miceli, 2024, private communication). Authors in \cite{2023A&A...680A..80G} have performed exploration of the variation of the peak energy for Si over the $(T,\tau)$ parameter space (their Fig.~B.1) and used the value $1.854$ keV that is very close to the value we have adopted.   

\end{document}